\documentclass[conference]{IEEEtran}
\IEEEoverridecommandlockouts

\usepackage{cite}
\usepackage{amsmath,amssymb,amsfonts}
\usepackage{graphicx}
\usepackage{textcomp}
\usepackage{xcolor}
\usepackage{pgfplots}
\pgfplotsset{compat=1.17}
\usetikzlibrary{arrows.meta,shapes.symbols,calc,patterns} 
\usepackage{url}
\usepackage{eso-pic}

\begin{document}
\AddToShipoutPictureFG*{\AtPageLowerLeft{\put(0,0){\makebox[\paperwidth][c]{\raisebox{0.30in}{\parbox{\textwidth}{\centering\scriptsize Accepted to the 2026 IEEE 1st Annual Integrated Sensing and Communication Conference (ISAC). \copyright~2026 IEEE. Personal use of this material is permitted. Permission from IEEE must be obtained for all other uses, in any current or future media, including reprinting/republishing this material for advertising or promotional purposes, creating new collective works, for resale or redistribution to servers or lists, or reuse of any copyrighted component of this work in other works.}}}}}}

\title{Fuse-then-Detect for Passive UAV Localization Using Multi-UE 5G Uplink Signals\\
\thanks{This work was supported in part by Samsung Research America.}}

\author{\IEEEauthorblockN{Wenyu Huang\IEEEauthorrefmark{1}, Nuria Gonz\'alez-Prelcic\IEEEauthorrefmark{1}, Vishnu Ratnam\IEEEauthorrefmark{2}, Murat Bayraktar\IEEEauthorrefmark{2}, and Charlie Jianzhong Zhang\IEEEauthorrefmark{2}} \IEEEauthorblockA{\IEEEauthorrefmark{1}Department of Electrical and Computer Engineering, University of California San Diego, USA \\
\IEEEauthorrefmark{2}Standards and Mobility Innovation Lab, Samsung Research America, USA \\
Email: \{wenyuhuang, ngprelcic\}@ucsd.edu, \{vishnu.r, m.bayraktar, jianzhong.z\}@samsung.com}
}
\IEEEaftertitletext{\vspace{-1.5\baselineskip}}

\maketitle

\begin{abstract}
Low-altitude uncrewed aerial vehicles (UAVs) can pose growing risks to airspace safety, security, and privacy.
Cellular infrastructure can passively sense them without dedicated radar hardware by exploiting integrated sensing and communication (ISAC) technology.
Most prior work exploits monostatic sensing or bistatic/multistatic configurations based on downlink measurements.
To the best of our knowledge, this paper presents the first uplink framework, where multiple user equipments (UEs) transmit sounding reference signal (SRS) pilots and the base station (BS) receives the UAV-scattered echoes.
Sensing from uplink SRS, however, introduces new challenges.
Each UE has its own oscillator and timing loop, so the channel estimate at the BS carries residual timing, frequency, and amplitude impairments that corrupt the UAV delay and Doppler.
Moreover, the UAV echo is weaker than both the line-of-sight (LOS) path and urban clutter, so detection from a single UE transmission is not reliable.
We address these challenges by designing a LOS-referenced synchronization scheme and a joint detector.
The synchronization reuses the existing timing advance (TA) command and an adjacent-occasion conjugate product to remove the residuals without additional signaling.
Then the detector searches a shared 3D state space and accumulates evidence across UEs.
It leverages a normalized contrast that exploits the bistatic geometry.
We evaluate the framework in a cluttered urban scene at frequency range 1 (FR1) with four pedestrian UEs and a $100$\,MHz 5G New Radio (NR) waveform.
The proposed pipeline achieves sub-nanosecond synchronization and a $4.84$\,m median 3D position error.
\end{abstract}


\section{Introduction}

Low-altitude uncrewed aerial vehicles (UAVs) pose growing challenges for airspace safety, critical-infrastructure security, and personal privacy. Conventional UAV sensing mainly relies on dedicated radar systems, e.g.,~\cite{fortem_trueview_r40}. In contrast, cellular ISAC reuses existing BSs as transmitters and/or receivers \cite{gonzalez2024integrated}, offering a lower-cost alternative that exploits the dense coverage of cellular networks.

Prior work on active and passive ISAC solutions for UAV detection has largely focused on monostatic configurations based on MIMO-OFDM radar~\cite{nuss2017mimo} or on bistatic/multistatic approaches that exploit downlink measurements, such as 5G passive radar~\cite{maksymiuk20235g} and SSB-based passive bistatic sensing~\cite{jopanya2025ssb}. Monostatic solutions require full-duplex hardware and careful self-interference management~\cite{BayraktarTWC2026}. Moreover, single-BS operation is vulnerable to obstructions and often performs poorly for distant UAVs~\cite{tang2025cooperative}. Downlink-based bistatic or multistatic sensing can benefit from the high BS transmit power but requires channel state information (CSI) feedback, depends on downlink scheduling and beam sweeping for target visibility~\cite{jopanya2025drone}, and remains vulnerable to blockage~\cite{zhao2025networked}. Multi-cell approaches can improve coverage and localization accuracy~\cite{liang2026uav}, but they require coordinated operation of multiple transmit and receive BSs, such as joint beamforming~\cite{zhang2025cooperative}, which substantially complicates network management, synchronization~\cite{yan2026asynchronous}, and calibration~\cite{Meng2025Cooperative}. Uplink sensing has received far less attention, and a concurrent uplink tracker assumes synchronized UEs and a target known to be present~\cite{he2026uplink}.

In this paper, we consider a distributed FR1 sensing architecture that exploits uplink transmissions. Specifically, we reuse standard SRS pilots, which are transmitted by multiple UEs for channel estimation~\cite{ts38211}. After the SRS transmissions are coarsely synchronized across UEs through the TA command~\cite{ts38213}, the UAV acts as a scatterer of these pilots, so the BS obtains multiple bistatic views of the target from different UE--UAV--BS paths. The approach requires no waveform modification and is compatible with existing infrastructure and current cellular operation for multi-user MIMO communication. However, uplink sensing also introduces several challenges.

To enable sensing as a byproduct of communication, we assume that the uplink channel estimates are obtained after conventional time and frequency synchronization for communication have been applied~\cite{wu2024sensing}. As a result, existing bistatic ISAC approaches that introduce a dedicated sensing synchronization stage after analog-to-digital conversion cannot be directly applied to our setting~\cite{pegoraro2024jump,ventura2026asymov}. Furthermore, typical FR1 SRS bandwidths provide only limited delay resolution, making it difficult to separate multipath components finely enough to support delay-domain time synchronization~\cite{ventura2026asymov}.

Detection under strong clutter is another key challenge. UEs transmit at approximately $23$\,dBm~\cite{ts38101}, far below typical BS downlink transmit powers, and the small physical size of UAVs further weakens the scattered echo, which is easily masked by urban clutter. A conventional \emph{detect-then-fuse} strategy~\cite{tang2025cooperative}, where each UE view is processed independently before fusion (e.g., threshold detection~\cite{liang2026uav} or feature extraction~\cite{yan2026asynchronous} at each station), is ineffective because no single UE is likely to isolate the UAV echo reliably.

To the best of our knowledge, this is the first system model and solution for passive UAV detection and localization using standard 5G NR uplink resources from multiple mobile UEs. The main contributions are as follows:
\begin{itemize}
    \item We formulate an uplink UAV sensing signal model under realistic 5G receiver impairments, including timing errors, Doppler-reference errors, gain variations, and phase errors across SRS occasions.
    \item We propose a LOS-referenced synchronization scheme that reuses the 5G NR TA command to suppress timing drift, phase drift, and gain variation for each UE without additional signaling.
    \item We develop a \emph{fuse-then-detect} pipeline that combines range-normalized evidence from multiple UEs in a 3D state space. By enforcing bistatic geometric consistency, the proposed method reveals UAV echoes that are too weak to be detected from any single UE view.
    \item We evaluate the method with a Friis bistatic channel and a 100 MHz NR waveform at FR1, where it outperforms recent multistatic baselines on the same channels.
\end{itemize}

\section{System Model}
\label{sec:system_model}

\subsection{Uplink SRS-Based UAV Sensing Scenario}
\label{subsec:scenario}

\begin{figure}[t]
    \centering
    \includegraphics[width=.85\columnwidth]{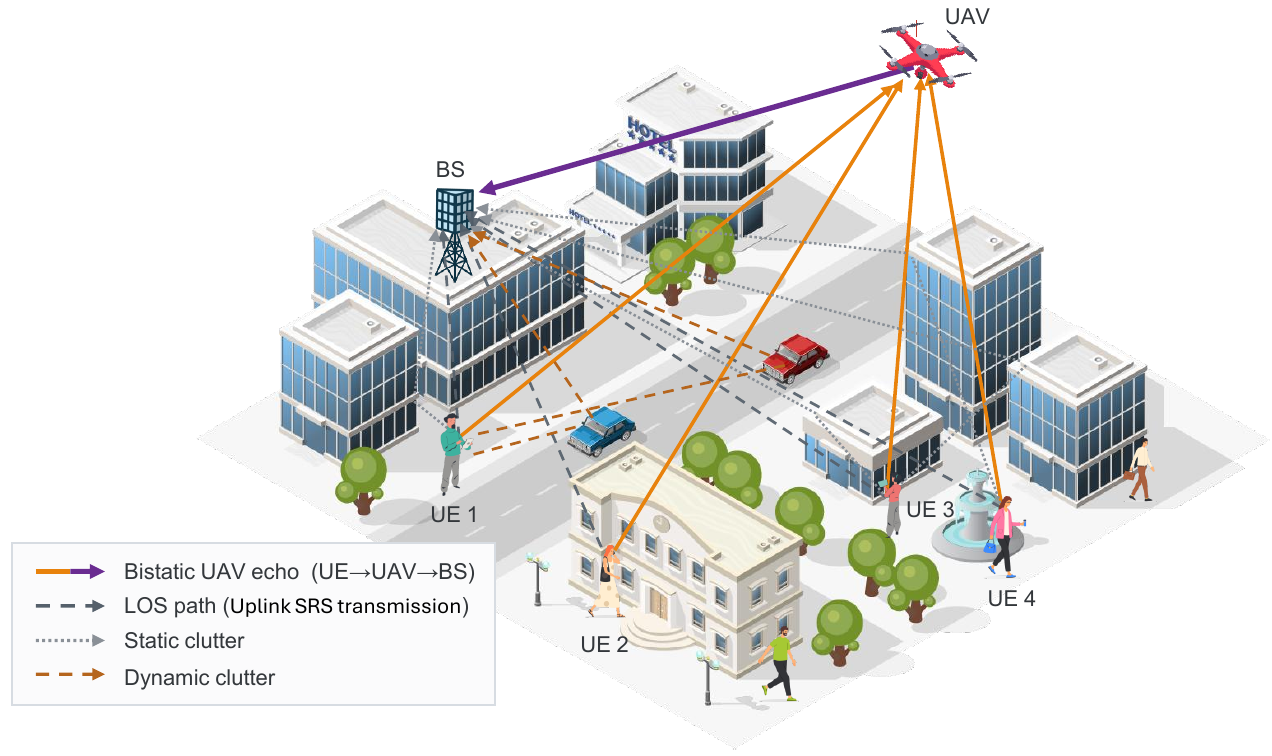}
    \caption{System model for urban uplink passive UAV sensing.}
    \label{fig:system_model}
\end{figure}

As shown in Fig.~\ref{fig:system_model}, we consider an urban single-cell uplink scenario, where $K$ ground UEs transmit standard 5G NR SRS to a BS with an $M_h\times M_v$ half-wavelength-spaced isotropic UPA, $M=M_hM_v$.
A small UAV acts as a passive aerial target and reflects the uplink SRS.
The propagation environment naturally includes static scatterers, such as buildings and ground reflections, as well as dynamic scatterers, such as ground vehicles.
The ground UEs move as pedestrians.
The system operates at carrier frequency $f_c$ with wavelength $\lambda=c/f_c$, subcarrier spacing $\Delta f$, and $N$ contiguous active subcarriers covering bandwidth $B=N\Delta f$.
For each SRS occasion $s=0,\ldots,S-1$, where $S$ is the number of
occasions in the observation window, the BS estimates the uplink channel from UE~$k$ on each active subcarrier, yielding $\mathbf{H}_k[n,s]\in\mathbb{C}^{M}$, $n\in\{0,\ldots,N-1\}$. The $S$ occasions are separated by $T_{\mathrm{SRS}}$. The full observation tensor $\widehat{\mathbf{H}}\in\mathbb{C}^{M\times K\times N\times S}$ serves as the input to the sensing algorithm.
Each occasion carries one SRS symbol.
The UEs use orthogonal SRS resources, i.e., distinct comb offsets or cyclic shifts~\cite{ts38211}, so the BS estimates their channels separately.
With a transmission comb $K_{\mathrm{TC}}$, the SRS occupies every $K_{\mathrm{TC}}$-th subcarrier, so it still spans $B$ and keeps the delay resolution, but the unambiguous delay shrinks to $1/(K_{\mathrm{TC}}\Delta f)$. For simplicity, $n$ indexes all $N$ subcarriers, each assumed to carry SRS.

\subsection{Uplink Receiver Processing and Residual Impairments}
\label{subsec:sync}

The BS performs standard 5G NR uplink synchronization and channel estimation before exporting $\widehat{\mathbf{H}}_k[n,s]$.
Time synchronization of the multi-UE SRS transmissions is performed through the TA procedure following~\cite{ts38213}.
Frequency synchronization at the BS first removes a coarse frequency offset using uplink reference signals.
The phase-tracking reference signal (PT-RS) is then used to track the residual common phase error~\cite{ts38211}.
Therefore, the channel estimate $\widehat{\mathbf{H}}_k[n,s]$ carries three classes of impairments, as described below, that are common to every propagation path of UE $k$ at occasion $s$.

\textit{Residual timing offset.}
To align transmission timing, the BS commands UE $k$ to advance its transmission by $T_{\mathrm{TA},k}[s]$, which approximates the true propagation delay
$\tau_k^{\mathrm{prop}}=\|\mathbf{p}_k-\mathbf{p}_{\mathrm{BS}}\|/c$, where $\mathbf{p}_k$ and $\mathbf{p}_{\mathrm{BS}}$ denote the positions of UE~$k$ and the BS.
The TA command differs from $\tau_k^{\mathrm{prop}}$ by an estimation error $\tau_{k,\mathrm{est}}[s]$ and a quantization error $\tau_{k,q}[s]$, yielding $T_{\mathrm{TA},k}[s]=\tau_k^{\mathrm{prop}}-\tau_{k,\mathrm{est}}[s]-\tau_{k,q}[s]$.
After this pre-compensation, the residual timing offset is
\begin{equation}
\Delta t_k[s]
= \tau_{k,\mathrm{est}}[s] + \tau_{k,q}[s]
+ \tau_{k,\mathrm{clk}}[s] + \delta t_k^{\mathrm{sync}}[s],
\label{eq:DeltaT}
\end{equation}
where $\tau_{k,\mathrm{clk}}[s]$ is the phase offset between UE $k$'s sampling clock and the BS timing reference.
Small updates of the receiver timing reference introduce occasion-to-occasion sampling jitter, denoted by $\delta t_k^{\mathrm{sync}}[s]$.

\textit{Common Frequency Reference.}
The frequency synchronization loop estimates and removes a Doppler reference $\nu_{\mathrm{ref},k}[s]$ for each UE from $\widehat{\mathbf{H}}_k$.
This reference absorbs the carrier frequency offset (CFO) and part of the dominant slow-time Doppler seen by the receiver.
Consequently, the Doppler of path $\ell$ in $\widehat{\mathbf{H}}_k$ takes the form
\begin{equation}
\nu_{k\ell}^{\mathrm{obs}}[s] = \bar\nu_{k\ell}[s] - \nu_{\mathrm{ref},k}[s],
\label{eq:Doppler}
\end{equation}
where $\bar\nu_{k\ell}[s]$ is the physical Doppler of path $\ell$.
Since $\nu_{\mathrm{ref},k}[s]$ has been absorbed by the receiver, it cannot be directly recovered from $\widehat{\mathbf{H}}_k$. Only $\nu_{k\ell}^{\mathrm{obs}}[s]$ are observable.

\textit{Common complex scalar.}
A complex scalar $\beta_k[s]\in\mathbb{C}$ summarizes the remaining common amplitude and phase distortion within occasion~$s$. It includes automatic gain control (AGC) variation, oscillator phase noise, residual common phase error after PT-RS tracking, and residual CFO~\cite{wu2024sensing,ratnam2024optimal}.

\subsection{Uplink Bistatic Channel Model}
\label{subsec:channel}
The estimated channel component of path $\ell$ from UE~$k$ is
\begin{align}
\mathbf{h}_{k\ell}[n,s]
&= \beta_k[s]\,\alpha_{k\ell}[s]\,
   e^{-j2\pi n\Delta f(\tau_{k\ell}[s]+\Delta t_k[s])} \nonumber\\
&\quad\cdot
   e^{\,j2\pi s T_{\mathrm{SRS}}\, \nu_{k\ell}^{\mathrm{obs}}[s]}\,
   \mathbf{a}_{\mathrm{BS}}(\phi_{k\ell}[s],\theta_{k\ell}[s]),
\label{eq:path}
\end{align}
where $\alpha_{k\ell}[s]\in\mathbb{C}$ is the complex path gain, $\tau_{k\ell}[s]$ is the absolute propagation delay of path $\ell$, and $\mathbf{a}_{\mathrm{BS}}(\phi,\theta)\in\mathbb{C}^M$ is the BS array response evaluated at the azimuth $\phi_{k\ell}[s]$ and elevation $\theta_{k\ell}[s]$ of the path arrival direction.

Assuming that the LOS path exists, we decompose the estimated channel into the LOS path, static clutter, dynamic clutter, the UAV echo, and residual noise as
\begin{equation}
\begin{aligned}
\widehat{\mathbf{H}}_k[n,s]
={}& \mathbf{H}_k^{\mathrm{LOS}}[n,s]
+ \mathbf{H}_k^{\mathrm{st}}[n,s]
+ \mathbf{H}_k^{\mathrm{dy}}[n,s] \\
&+ \mathbf{H}_k^{\mathrm{UAV}}[n,s]
+ \mathbf{E}_k[n,s].
\end{aligned}
\label{eq:Hsplit}
\end{equation}
Here, $\mathbf{H}_k^{\mathrm{LOS}}[n,s]$ is the direct UE--BS path, $\mathbf{H}_k^{\mathrm{st}}[n,s]$ represents static clutter from buildings and ground reflections, $\mathbf{H}_k^{\mathrm{dy}}[n,s]$ represents dynamic clutter from ground vehicles, and $\mathbf{H}_k^{\mathrm{UAV}}[n,s]$ is the UE--UAV--BS bistatic echo. $\mathbf{E}_k[n,s]$ collects channel estimation noise.
For the UAV echo, with UAV position $\mathbf{q}[s]$, the propagation delay is
\begin{equation}
\tau_k^{\mathrm{UAV}}[s]
= \frac{
\|\mathbf{p}_{\mathrm{UE},k}[s]-\mathbf{q}[s]\|
+\|\mathbf{q}[s]-\mathbf{p}_{\mathrm{BS}}\|
}{c}.
\label{eq:tauUAV}
\end{equation}
UE motion also contributes to the observed bistatic Doppler.

\section{Proposed Uplink UAV Sensing Framework}
\label{sec:pipeline}
As summarized in Fig.~\ref{fig:pipeline}, our framework synchronizes each UE channel and suppresses the clutter, fuses the evidence of all UEs in a shared 3D state space, and then validates and localizes the UAV with quantities already available at the BS.

\begin{figure}[t]
\centering
\scalebox{0.93}{
\begingroup
\definecolor{cblue}{RGB}{70,130,200}
\definecolor{cpurple}{RGB}{120,80,170}
\definecolor{corange}{RGB}{225,130,60}
\definecolor{clink}{RGB}{35,85,175}
\newcommand{\secref}[1]{{\color{clink}\S\,#1}}
\begin{tikzpicture}[
  x=1pt,y=1pt,
  font=\footnotesize,
  >={Stealth[length=4pt,width=3.5pt]},
  panel/.style={draw=#1!70!black, fill=#1!10, line width=0.6pt, rounded corners=2pt},
  ghost/.style={draw=#1!55!black, fill=#1!6, line width=0.5pt, rounded corners=2pt},
  blk/.style={draw=#1!65!black, fill=#1!22, line width=0.5pt, rounded corners=1.5pt,
              inner xsep=2.5pt, inner ysep=2pt, align=center},
  chev/.style={signal, signal to=east, signal from=west, draw=cblue!65!black, fill=cblue!22,
               line width=0.5pt, inner xsep=3.5pt, inner ysep=2.2pt, align=center},
  flow/.style={->, line width=0.6pt, draw=black!75},
  gate/.style={->, densely dashed, line width=0.55pt, draw=cpurple},
  step/.style={circle, fill=#1, text=white, inner sep=0.4pt, minimum size=9pt,
               font=\footnotesize\bfseries},
  ttl/.style={inner sep=1.5pt, font=\footnotesize\bfseries\itshape},
  drone/.pic={\draw[black!75, line width=0.6pt] (-5,3) -- (5,-3) (-5,-3) -- (5,3);
              \fill[black!75] (0,0) circle (2);
              \foreach \p in {(-5,3),(5,3),(-5,-3),(5,-3)} \draw[black!75, fill=corange!60, line width=0.4pt] \p circle (1.9);},
]
\draw[ghost=cblue] (5,5) rectangle (250.5,-73);
\draw[ghost=cblue] (2.5,2.5) rectangle (248,-75.5);
\draw[panel=cblue] (0,0) rectangle (245.5,-78);
\node[ttl, text=cblue!70!black, anchor=west] at (3,-7) {\underline{Each UE}};
\node[anchor=north east, inner sep=1.5pt] at (245,-1) {$k=1,\dots,K$ in parallel};

\node[align=left, inner sep=1pt, anchor=west] (in) at (3,-29)
     {$\widehat{\mathbf H}_k[n,s]$\\[-1pt]$T_{\mathrm{TA},k}$};
\node[blk=cblue, anchor=west, text width=92pt] (sync) at (44,-29)
     {\textbf{LOS-ref.\ sync} \secref{III-A}\\[-1pt]
      TA $\to$ LOS ref.\ $g_k$\\[0.5pt]
      occasion product $Q_k$};
\node[blk=cblue, anchor=west, text width=78pt] (clu) at (156,-29)
     {\textbf{clutter supp.} \secref{III-B}\\[-1pt]
      static mean\\[-1pt]
      low-elev.\ projector $\mathbf F$};
\draw[flow] (in.east) -- (sync.west);
\draw[flow] (sync.east) -- (clu.west)
   node[pos=0.5, above=0.5pt, inner sep=0.5pt] {$\widetilde{\mathbf H}_k$};

\node[chev, anchor=west] (c1) at (11,-65) {\secref{III-C}\; beamform to $\boldsymbol\theta$};
\node[chev, anchor=west] (c2) at ([xshift=-3pt]c1.east) {contrast $\Lambda_k$};
\node[chev, anchor=west] (c3) at ([xshift=-3pt]c2.east) {range-norm.\ $\zeta_k(\boldsymbol\theta)$};
\draw[flow] (clu.south) -- ++(0,-8.5) -| ([xshift=8pt]c1.north west);
\node[fill=cblue!10, inner sep=1pt] at (118,-51.0) {$\widetilde{\mathbf H}^{(\mathrm{res})}_k[n,s]$};

\draw[panel=cpurple] (0,-90) rectangle (245.5,-136.5);
\node[ttl, text=cpurple!80!black, anchor=west] at (3,-98) {\underline{Fuse before detect}};
\node[anchor=north west, align=left, inner sep=1pt, text width=118pt] at (4,-107)
  {shared 3-D grid $\boldsymbol\theta=(az,el,\rho)$\\[-1pt]
   trimmed mean $T(\boldsymbol\theta)$, $\ge K_{\min}$ UEs};
\begin{scope}[shift={(184,-127)}]  
  \def\a{24}\def\dx{8}\def\dy{6}\def\n{3}
  \fill[cpurple!12] (0,0) rectangle (\a,\a);
  \fill[cpurple!22] (0,\a) -- (\a,\a) -- (\a+\dx,\a+\dy) -- (\dx,\a+\dy) -- cycle;
  \fill[cpurple!32] (\a,0) -- (\a+\dx,\dy) -- (\a+\dx,\a+\dy) -- (\a,\a) -- cycle;
  \fill[corange!35] (2*\a/\n-1,\a/\n-1) rectangle (3*\a/\n+1,2*\a/\n+1);
  \fill[corange!90] (2*\a/\n,\a/\n) rectangle (3*\a/\n,2*\a/\n);
  \foreach \i in {0,...,\n} {
    \draw[cpurple!60!black, line width=0.3pt] (\i*\a/\n,0) -- (\i*\a/\n,\a) -- (\i*\a/\n+\dx,\a+\dy);
    \draw[cpurple!60!black, line width=0.3pt] (0,\i*\a/\n) -- (\a,\i*\a/\n) -- (\a+\dx,\i*\a/\n+\dy);
    \draw[cpurple!60!black, line width=0.3pt] (\i*\dx/\n,\a+\i*\dy/\n) -- (\a+\i*\dx/\n,\a+\i*\dy/\n);
    \draw[cpurple!60!black, line width=0.3pt] (\a+\i*\dx/\n,\i*\dy/\n) -- (\a+\i*\dx/\n,\a+\i*\dy/\n);
  }
  \draw[cpurple!70!black, line width=0.5pt] (0,0) rectangle (\a,\a);
  \draw[cpurple!70!black, line width=0.5pt] (0,\a) -- (\dx,\a+\dy) -- (\a+\dx,\a+\dy) -- (\a+\dx,\dy) -- (\a,0);
  \draw[cpurple!70!black, line width=0.5pt] (\a,\a) -- (\a+\dx,\a+\dy);
  \node[anchor=north, inner sep=0.5pt] at (\a/3,-0.5) {$az$};
  \node[anchor=east, inner sep=0.5pt] at (-0.5,\a/2) {$el$};
  \node[anchor=west, inner sep=0.5pt] at (\a+\dx+0.5,\dy/2+1) {$\rho$};
\end{scope}
\foreach \dx/\t in {-12/194,-4/200,4/206,12/212} {
  \draw[flow] ([xshift=\dx pt]c3.south) to[out=-90,in=90] (\t,-100);
}
\node[anchor=east, inner sep=1pt] at ([xshift=-16pt]c3.south |- {(0,-84.5)}) {$\zeta_k(\boldsymbol\theta)$};
\draw[gate] (184,-104) -- (134,-104) -- (c2.south -| {(134,0)});
\node[anchor=east, inner sep=1pt, text=cpurple!90!black] at (131,-84.5) {gate $\tau_k^{\star}(\boldsymbol\theta)$};

\draw[panel=corange] (0,-148) rectangle (245.5,-191.5);
\node[ttl, text=corange!75!black, anchor=west] at (3,-154.5) {\underline{Across UEs}};
\node[blk=corange, anchor=west, text width=98pt] (det) at (4,-175)
  {\textbf{detect} \secref{III-C}\\[-1pt]
   local refinement\\[-1pt]Doppler and spatial checks};
\node[blk=corange, anchor=west, text width=60pt] (loc) at (140,-175)
  {\textbf{locate} \secref{III-D}\\[-1pt]
   closed-form\\[-1pt]ray--delay solve};
\pic at (228,-170) {drone};
\node[anchor=north, inner sep=1pt] at (228,-177) {$\hat{\mathbf q}_{\mathrm{UAV}}$};
\draw[flow] (det.east) -- (loc.west)
  node[pos=0.5, above=1pt, inner sep=0.5pt] {$\hat{\mathbf u}$, $\Delta\hat\tau_k$};
\draw[flow] (loc.east) -- (219,-175);
\draw[flow] (205,-127) -- (205,-142.3) -| (det.north)
  node[pos=0.45, fill=white, inner xsep=1pt, inner ysep=0.3pt, font=\scriptsize] {$\hat{\boldsymbol\theta}=\arg\max_{\boldsymbol\theta} T(\boldsymbol\theta)$};
\end{tikzpicture}
\endgroup%
}%
\caption{Processing flow of the proposed framework.}
\label{fig:pipeline}
\end{figure}

\subsection{LOS-Referenced Synchronization}
\label{subsec:sync_alg}

In Eq.~\eqref{eq:path}, $\beta_k[s]$ and the receiver Doppler reference $\nu_{\mathrm{ref},k}$ corrupt the delay and Doppler signatures needed for sensing.
A reference path is needed to remove these distortions.
We observe that the LOS path is a natural choice.
It is usually the strongest path, shares the same common distortion $\beta_k[s]$ with the UAV echo, and has an approximate delay already provided by the TA command.
Therefore, assuming that a UE--BS LOS path is available for each UE, we reuse only the TA command, requiring no additional signaling or UE modification.
This assumption can be relaxed.
Since $\Delta t_k[s]$, $\nu_{\mathrm{ref},k}[s]$, and $\beta_k[s]$ are common to all paths of UE~$k$, any stable dominant path can be the reference.
If the LOS is blocked, the reference is a strong static reflection. The synchronization is unchanged, but every referenced delay of that UE shifts by the excess path length of the reflection (see Section~\ref{subsec:nlos}).

\textit{Step 1: Estimate LOS delay and direction.}
The TA command $T_{\mathrm{TA},k}$ provides a coarse LOS delay $\hat\tau_{k,\mathrm{LOS}}=T_{\mathrm{TA},k}$.
This delay is refined by searching a narrow window around the TA prediction and accumulating power over the first $S_0$ SRS occasions, with $S_0 < S$.
At the refined delay bin, the BS forms a spatial covariance matrix, and the LOS steering vector $\hat{\mathbf{a}}_{k,\mathrm{LOS}}$ is taken as its dominant eigenvector.
The covariance is invariant to the unknown phase $\angle\beta_k[s]$, so this direction estimate is robust to phase variations across SRS occasions.

\textit{Step 2: Construct the LOS reference.}
We remove the predicted LOS delay and project the channel onto $\hat{\mathbf{a}}_{k,\mathrm{LOS}}$
\begin{equation}
g_k[n,s]
= \hat{\mathbf{a}}_{k,\mathrm{LOS}}^H
\widehat{\mathbf{H}}_k[n,s]
e^{j2\pi n\Delta f\,\hat\tau_{k,\mathrm{LOS}}}.
\label{eq:yref}
\end{equation}
This operation replaces the delay term $\tau_{k,\mathrm{LOS}}[s]+\Delta t_k[s]$ by $\epsilon_{k,\tau}^{\mathrm{LOS}}[s]+\Delta t_k[s]$, where $\epsilon_{k,\tau}^{\mathrm{LOS}}[s]=\tau_{k,\mathrm{LOS}}[s]-\hat\tau_{k,\mathrm{LOS}}[s]$ is small.
Thus, $g_k[n,s]$ is a LOS-centered reference whose residual phase mainly contains $\Delta t_k[s]$ and $\beta_k[s]$.

\textit{Step 3: Remove the unknown phase terms by adjacent-occasion product.}
The timing offset $\Delta t_k[s]$ adds a phase slope across $n$. At a single occasion, $\angle\beta_k[s]$, the LOS path phase, and the LOS slow-time Doppler all add a flat phase and cannot be separated, but sensing only needs their sum removed.
The adjacent-occasion conjugate product
\begin{equation}
Q_k[n,s] = g_k[n,s]\cdot g_k[n,s-1]^{*}
\label{eq:Qproduct}
\end{equation}
cancels the LOS path phase, which is constant across occasions, and exposes only the inter-occasion increments
\begin{equation}
Q_k[n,s] \approx A_k[s]\,e^{j\psi_k[s]}\,e^{-j2\pi n\Delta f\,\eta_k[s]},
\label{eq:Qmodel}
\end{equation}
where $\eta_k[s]{=}\Delta t_k[s]{-}\Delta t_k[s{-}1]$ is the timing increment, $\psi_k[s]$ absorbs the LOS Doppler step and $\angle\beta_k[s]{-}\angle\beta_k[s{-}1]$, and $A_k[s]{\ge}0$.
In $Q_k$, $\eta_k[s]$ appears as the slope phase across $n$ and $\psi_k[s]$ as its constant component.
The TA error and the fixed timing bias of the UE are also constant and cancel, so only small increments are estimated and the absolute LOS delay and phase never need to be known.

\textit{Step 4: Increment estimation and cumulative correction.}
Both increments are estimated from $Q_k$
\begin{equation}
\begin{aligned}
\hat\eta_k[s]&=-\tfrac{1}{2\pi\Delta f}\angle\big(\textstyle\sum_{n}Q_k[n{+}1,s]\,Q_k^*[n,s]\big),\\
\hat\psi_k[s]&=\angle\big(\textstyle\sum_{n}Q_k[n,s]\,e^{j2\pi n\Delta f\hat\eta_k[s]}\big).
\end{aligned}
\label{eq:eta_est}
\end{equation}
Then the cumulative timing and phase corrections are $\hat T_k[s]=\sum_{u=1}^{s}\hat\eta_k[u]$ and $\hat\Psi_k[s]=\sum_{u=1}^{s}\hat\psi_k[u]$.
To first order, the increment errors cancel in the sum, so the error of $\hat T_k[s]$ depends only on occasions $s$ and $0$. Second order terms of order $1/(\sqrt{N}\,\mathrm{SNR}_{\mathrm{LOS}})$ add up like a random walk but stay small within a short window. The adjacent product assumes that every occasion is valid. If an occasion is faded, missing, or irregularly spaced, its increment is wrong, and the error enters all later corrections. A fixed reference occasion $s_0$ avoids this: $Q_k[n,s]=g_k[n,s]\,g_k^{*}[n,s_0]$ gives the total change since $s_0$ directly, without summing increments, so a bad occasion affects only itself. The corrected channel then differs only by a constant timing and phase offset.
The amplitude $\hat r_k[s]$ is the subcarrier median of $|\widehat{\mathbf a}_{k,\mathrm{LOS}}^{H}\widehat{\mathbf H}_k[n,s]|$, normalized to its $s=0$ value.
The LOS projection isolates the dominant path, so $\hat r_k[s]$ captures the SRS occasion gain $|\beta_k[s]|$ up to a constant scale. The synchronized channel is
{\small\setlength{\baselineskip}{12pt}
\begin{equation}
\begin{aligned}
\widetilde{\mathbf{H}}_k[n,s]
&=
\frac{
\widehat{\mathbf{H}}_k[n,s]
e^{j2\pi n\Delta f(\hat\tau_{k,\mathrm{LOS}}+\hat T_k[s])-j\hat\Psi_k[s]}
}{\hat r_k[s]} \\
&\approx
\sum_{\ell\in\mathcal{P}_k}
\widetilde\alpha_{k\ell}
e^{-j2\pi n\Delta f\widetilde\tau_{k\ell}}
e^{j2\pi sT_{\mathrm{SRS}}\Delta\nu_{k\ell}}
\mathbf{a}_{k\ell}
+\widetilde{\mathbf{W}}_k[n,s],
\end{aligned}
\label{eq:Hsync}
\end{equation}
}%
where $\widetilde\tau_{k\ell} = \tau_{k\ell}-\hat\tau_{k,\mathrm{LOS}}+b_{k,\tau}$ is the LOS-referenced delay, $\Delta\nu_{k\ell} = \bar\nu_{k\ell}-\bar\nu_{k,\mathrm{LOS}}+b_{k,\nu}$ is the LOS-referenced differential Doppler, $\widetilde\alpha_{k\ell}$ is the normalized path gain, and $\widetilde{\mathbf{W}}_k[n,s]$ collects residual estimation errors.
The small residual biases $b_{k,\tau}$ and $b_{k,\nu}$ are absorbed by the geometric gates.

\subsection{Clutter Suppression}
\label{subsec:clutter}
We apply two filters that exploit physical signatures of the clutter. Static scatterers from buildings and ground reflections are nearly invariant in slow time, since pedestrian UE motion induces negligible delay variation within the observation window.
Ground vehicles and other near-ground dynamic scatterers usually arrive at the BS from low elevation angles, which makes them separable.
The first filter removes the static component by subtracting the slow-time sample mean
\begin{equation}
\widetilde{\mathbf{H}}_k^{(1)}[n,s] = \widetilde{\mathbf{H}}_k[n,s] - \frac{1}{S}\sum_{s'=0}^{S-1}\widetilde{\mathbf{H}}_k[n,s'].
\label{eq:static_sub}
\end{equation}
The second filter exploits the elevation. A steering dictionary $\mathbf{A}_{\mathrm{low}}$ spanning the elevation range $[\theta_{\mathrm{low}}, \theta_{\mathrm{high}}]$ forms a soft spatial projector $\mathbf{F} = \mathbf{I}_M - \xi\mathbf{P}_{\mathrm{low}}$ with weight $\xi$, where $\mathbf{P}_{\mathrm{low}}$ is the projection onto the column space of $\mathbf{A}_{\mathrm{low}}$. Applying $\mathbf{F}$ to the antenna axis of $\widetilde{\mathbf{H}}_k^{(1)}$ produces the clutter-suppressed residual $\widetilde{\mathbf{H}}_k^{(\mathrm{res})}[n,s]$.

\subsection{Geometry-Coupled Bistatic Fusion}
\label{subsec:fusion}

A true UAV echo is consistent with one 3D position across UEs, whereas clutter generally is not.
We therefore search a shared state space and accumulate weak evidence across UEs.
Within the short observation window, we write the UE position as $\mathbf{p}_{\mathrm{UE},k}$ for brevity.
Let $\boldsymbol{\theta}=(az,el,\rho)$ denote a candidate UAV state, with candidate position
$\hat{\mathbf{q}}(\boldsymbol{\theta})=\mathbf{p}_{\mathrm{BS}}+\rho\hat{\mathbf{u}}(az,el)$.
For UE~$k$, define $d_k=\|\mathbf{p}_{\mathrm{UE},k}-\mathbf{p}_{\mathrm{BS}}\|$.
The predicted LOS-referenced bistatic excess delay is
\begin{equation}
\tau_k^{\star}(\boldsymbol{\theta})
=
\frac{
\|\mathbf{p}_{\mathrm{UE},k}-\hat{\mathbf{q}}(\boldsymbol{\theta})\|
+\rho
-d_k
}{c}.
\label{eq:tau_pred}
\end{equation}

\textit{Evidence extraction.}
For candidate $\boldsymbol{\theta}$, we beamform the residual channel toward the candidate direction $\hat{\mathbf{u}}(az,el)$
\begin{equation}
X_k[n,s;\boldsymbol{\theta}]
= \mathbf{a}_{\mathrm{BS}}^{H}(\boldsymbol{\theta})
\widetilde{\mathbf{H}}_k^{(\mathrm{res})}[n,s].
\label{eq:beam}
\end{equation}
Then applying an IFFT over $n$ and an FFT over $s$ gives the delay-Doppler response $X_k(\tau,\nu;\boldsymbol{\theta})$ and power map
$P_k(\tau,\nu;\boldsymbol{\theta})=|X_k(\tau,\nu;\boldsymbol{\theta})|^2$.
Within the predicted gate
$\Omega_k(\boldsymbol{\theta})=\{(\tau,\nu)\,|\,|\tau-\tau_k^{\star}|\le\Delta_\tau,\;|\nu|\le\nu_{\mathrm{N}}\}$,
UE~$k$ compares the strongest gated response with a local background level
\begin{equation}
\Lambda_k(\boldsymbol{\theta})
= \log\!\!\max_{(\tau,\nu)\in\Omega_k}\!\! P_k(\tau,\nu;\boldsymbol{\theta})
- \log F_k(\boldsymbol{\theta}),
\label{eq:Lambda}
\end{equation}
where $F_k(\boldsymbol{\theta})$ is the $p_0$ percentile of the delay-max profile of $P_k$ outside $\Omega_k$.
This log contrast measures how much the candidate stands out from its local background, which reduces the effect of link-budget differences and of raw clutter power.

\textit{Contrast normalization.}
A strong clutter path aligned with $\hat{\mathbf{u}}$ still gives a large $\Lambda_k$ over many candidate ranges, whereas a UAV echo stands out only near its range $\rho^{\star}$.
We therefore normalize $\Lambda_k$ along the range axis at each direction $(az,el)$
\begin{equation}
\zeta_k(\boldsymbol{\theta}) =
\frac{\Lambda_k(\boldsymbol{\theta})-\operatorname{med}_{\rho}\Lambda_k}
{\operatorname{med}_{\rho}\left|\Lambda_k-\operatorname{med}_{\rho}\Lambda_k\right|}.
\label{eq:zeta}
\end{equation}
So flat clutter gives $\zeta_k\!\approx\!0$ while a localized UAV peak gives a large positive value near $\rho^\star$.

\textit{Fusion across UEs.} Let $\zeta_{[1]}\ge\cdots\ge\zeta_{[K]}$ be the order of $\{\zeta_k(\boldsymbol{\theta})\}$. The fused detection score is the trimmed mean
\begin{equation}
T(\boldsymbol{\theta}) = \frac{1}{K-1}\sum_{k=1}^{K-1} \zeta_{[k]}(\boldsymbol{\theta}),
\label{eq:fusion}
\end{equation}
which discards the weakest UE so that one blocked or unfavorable path does not veto detection. Let $K_{\min}$ denote the minimum number of supporting UEs.
Candidates are penalized if fewer than $K_{\min}$ UEs have $\zeta_k>\zeta_{\min}$.
Such weak cross-UE support is more likely caused by sidelobe or clutter alignment.
\begin{figure}[t]
\centering
\begingroup
\definecolor{cOur}{RGB}{204,128,82}
\definecolor{cBlue}{RGB}{88,126,168}
\definecolor{cGrn}{RGB}{92,146,128}
\definecolor{cPur}{RGB}{138,114,166}
\definecolor{cPink}{RGB}{186,104,138}
\pgfplotsset{
  zbase/.style={
    scale only axis, height=1.95cm,
    tick label style={font=\scriptsize}, label style={font=\scriptsize},
    axis line style={black!40, line width=0.4pt}, axis lines=box,
    tick pos=left, tick align=outside, major tick length=1.8pt, every tick/.style={black!40},
    grid=major, grid style={black!12, dashed, line width=0.3pt},
    xmin=45, xmax=200, xtick={60,100,140,180}, minor xtick={80,120,160},
    xlabel style={yshift=3pt}, ylabel style={yshift=-5pt},
    clip=true,
  },
}
\begin{tikzpicture}[font=\scriptsize]
\begin{axis}[
  name=A, zbase, width=4.1cm, ymin=-2.5, ymax=15.5, ytick={0,5,10,15},
  xlabel={(a) Candidate range $\rho$ [m]}, ylabel={$\zeta_k(\rho)$},
  legend style={at={(0.97,0.97)}, anchor=north east, font=\fontsize{6}{6.5}\selectfont,
    draw=black!30, fill=white, fill opacity=0.75, draw opacity=1, text opacity=1, sharp corners,
    legend columns=2, column sep=3pt, inner xsep=1.5pt, inner ysep=1pt, row sep=-2pt},
  legend image post style={draw opacity=1},
  legend image code/.code={\draw[#1] (0cm,0cm) -- (0.26cm,0cm);},
]
\draw[cPink!80, densely dashed, line width=0.5pt] (axis cs:134.8,-2.5) -- (axis cs:134.8,15.5);
\addplot[cBlue, line width=0.8pt] table[x=x,y=y] {figure/fig3_ue1.dat};
\addplot[cOur,  line width=0.8pt] table[x=x,y=y] {figure/fig3_ue2.dat};
\addplot[cGrn,  line width=0.8pt] table[x=x,y=y] {figure/fig3_ue3.dat};
\addplot[cPur,  line width=0.8pt] table[x=x,y=y] {figure/fig3_ue4.dat};
\legend{UE\,1, UE\,2, UE\,3, UE\,4}
\node[anchor=east, text=cPink, font=\scriptsize, inner sep=1.5pt] at (axis cs:134.8,7.4) {$\rho^{\star}$};
\end{axis}
\begin{axis}[
  name=B, at={(A.east)}, anchor=west, xshift=22pt,
  zbase, width=2.9cm, ymin=-0.5, ymax=6.8, ytick={0,2,4,6},
  xlabel={(b) Candidate range $\rho$ [m]}, ylabel={$T(\rho)$},
]
\addplot[draw=none, fill=cBlue!15, forget plot] table[x=x,y=y] {figure/fig3_T.dat} \closedcycle;
\draw[cPink!80, densely dashed, line width=0.5pt] (axis cs:134.8,-0.5) -- (axis cs:134.8,6.8);
\addplot[cBlue!85!black, line width=0.8pt] table[x=x,y=y] {figure/fig3_T.dat};
\addplot[only marks, mark=star, mark size=2.6pt, cOur, line width=0.7pt] coordinates {(134,5.87)};
\node[anchor=north east, text=cOur!85!black, font=\scriptsize, inner sep=1.5pt, align=right] at (axis cs:132,5.6) {$\hat\rho{=}134$\,m};
\end{axis}
\end{tikzpicture}%
\endgroup%
%
\caption{(a) $\zeta_k(\rho)$ at the UAV direction for each UE and (b) the fused score $T(\rho)$. The dashed line marks the true range $\rho^{\star}$.}
\label{fig:zeta_fusion}
\end{figure}
Fig.~\ref{fig:zeta_fusion} illustrates a fused peak at $\hat\rho=134$~m supported by UEs~1--3 but not by UE~4.

\textit{Local refinement.}
We refine the top-ranked candidates by a local search around the coarse grid maximum using the same $\zeta_k$ criterion.
The contrast in Eq.~\eqref{eq:Lambda} is computed within a narrow Doppler range around the coarse peak to reduce inconsistent clutter.
Parabolic interpolation around the refined peak then gives an off-grid range estimate.

\textit{Cluster validation.}
Each refined candidate is finally checked against two physical constraints.
We require the supporting UEs' wrapped Doppler peaks to fit one bounded UAV velocity, allowing for pedestrian UE motion, and their angle peaks to share one BS arrival direction.

\subsection{UAV Position Estimation}
\label{subsec:position}
After cluster validation, we obtain the BS-side UAV direction $\hat{\mathbf{u}}$ and the LOS-referenced excess delay $\Delta\hat{\tau}_k$ for each supporting UE.
The UAV is then constrained to lie on the ray $\hat{\mathbf{q}}=\mathbf{p}_{\mathrm{BS}}+\rho\hat{\mathbf{u}}$.
Substituting this ray into the bistatic delay constraint
$c\Delta\hat{\tau}_k=\|\hat{\mathbf{q}}-\mathbf{p}_{\mathrm{UE},k}\|+\rho-d_k$
gives
\begin{equation}
\hat\rho_k =
\frac{c\Delta\hat\tau_k\,(c\Delta\hat\tau_k+2d_k)}
{2\bigl(\mathbf{d}_k^{\top}\hat{\mathbf{u}}+c\Delta\hat\tau_k+d_k\bigr)},
\label{eq:rho_closed}
\end{equation}
where $\mathbf{d}_k=\mathbf{p}_{\mathrm{BS}}-\mathbf{p}_{\mathrm{UE},k}$ and $d_k=\|\mathbf{d}_k\|$.
After removing outliers, we average the remaining $\hat\rho_k$ values to obtain $\hat\rho$.
The UAV position estimate is
$\hat{\mathbf{q}}_{\mathrm{UAV}}=\mathbf{p}_{\mathrm{BS}}+\hat\rho\hat{\mathbf{u}}$.

\section{Simulation Results}
\label{sec:simulation}

We simulate a cluttered urban uplink scenario with the parameters in Table~\ref{tab:sim_params}. The four pedestrian UEs start at $[80,50]$, $[100,-60]$, $[-80,60]$, and $[-100,-50]$\,m around the BS. This layout surrounds the BS and favors the fusion. In a real network, the scheduler decides which UEs transmit SRS, and they may cluster on one side of the BS, where their delay constraints are less diverse. The $140$ trials cover $14$ static-clutter realizations with $10$ trials each, and the sensing algorithm only knows the pedestrian speed bound of $1$\,m/s.
A declaration within $10$\,m of the UAV is useful. With the standard comb $K_{\mathrm{TC}}=4$, the four UEs share one SRS symbol in each period, about $0.36\%$ of the resource elements, and the channels are simulated on all $N$ subcarriers.
The UAV echo follows the bistatic Friis radar equation, $|\alpha_{k,\mathrm{UAV}}|=\lambda\sqrt{\sigma_{\mathrm{UAV}}}/\big((4\pi)^{3/2}\|\mathbf{p}_{\mathrm{UE},k}-\mathbf{q}_0\|\,\|\mathbf{q}_0-\mathbf{p}_{\mathrm{BS}}\|\big)$, with initial UAV position $\mathbf{q}_0$ and RCS $\sigma_{\mathrm{UAV}}$, which places the echo $44$ to $49$\,dB below the LOS. The amplitude is fixed over the short window, while the UAV delay and Doppler are updated across SRS occasions. RCS fluctuation and rotor micro-Doppler are not considered.
Receiver impairments after coarse synchronization are injected for each UE~$k$.
A timing shift $\Delta t_k[s]=b_{t,k}+\epsilon_{t,k}[s]$ produces the subcarrier phase ramp $e^{-j2\pi n\Delta f\,\Delta t_k[s]}$, where the fixed bias of each UE $b_{t,k}\!\sim\!\mathcal{U}(\pm130\,\mathrm{ns})$ and the jitter at each occasion $\epsilon_{t,k}[s]\!\sim\!\mathcal{N}(0,15^2\,\mathrm{ns}^2)$ model the half-step quantization error of a $30$~kHz TA command (step $512T_c\!\approx\!260$~ns)~\cite{ts38213}.
A residual common Doppler $\epsilon_{\nu,k}[s]\!\sim\!\mathcal{U}(\pm30\,\mathrm{Hz})$ is applied to every path of UE~$k$, modeling the CFO residual
after BS coarse-loop removal.
The common complex scalar is modeled as $\beta_k[s]=10^{\gamma_k[s]/20}e^{j\angle\beta_k[s]}$, where $\gamma_k[s]$ includes slow gain drift, gain jitter, and AGC jumps with probability $0.06$ and clipping at $\pm8$~dB.
The phase $\angle\beta_k[s]$$\sim\mathcal{N}(0,(3^\circ)^2)$ models residual common phase error.

\begin{table}[t]
\centering
\caption{Main simulation parameters.}
\label{tab:sim_params}
\scriptsize
\setlength{\tabcolsep}{1.6pt}
\renewcommand{\arraystretch}{1.02}
\begin{tabular}{@{}l c l c@{}}
\hline
Param. & Value & Param. & Value \\
\hline
Carrier freq. $f_c$ & $3.5$ GHz
& SCS $\Delta f$ & $30$ kHz \\
Subcarriers $N$ & $3276$
& SRS comb $K_{\mathrm{TC}}$ & $4$ \\
SRS period $T_{\mathrm{SRS}}$ & $10$ ms
& SRS occasions $S$ & $50$ \\
BS array & $4\times4$ UPA
& UE number $K$ & $4$ \\
\hline
BS position & $[0,0,25]$ m
& UE height & $1.5$ m \\
UAV start & $[40,0,60]$ m
& UAV velocity & $[0,8,0]$ m/s \\
UAV RCS & $0$ dBsm
& MC trials & $140$ \\
Static scatterers & $50$ within $30$--$200$ m
& Building refl. & $4$, $45$--$95$ m tall \\
Vehicles & $3$ at $5$--$14$ m/s
& UE speeds & $0.75$--$0.95$ m/s \\
\hline
Low-elev. projector & $[-45,-12]^\circ$, $\xi{=}0.7$
& LOS init. $S_0$ & $10$ \\
Az. grid & $[-180,180]^\circ/6^\circ$
& El. grid & $[-15,75]^\circ/4^\circ$ \\
Range grid & $[45,200]$ m / $5$ m
& Range refine & $\pm8$ m \\
Coarse delay gate $\Delta_\tau$ & $45$ ns
& Refined delay gate & $35$ ns \\
Doppler gate $\nu_{\mathrm N}$ & $50$ Hz
& Bg. percentile $p_0$ & $60$th \\
UAV speed range & $[1.5,30]$ m/s
& AoA agreement & within $35^\circ$ \\
Min. support $K_{\min}$ & $3$
& Support thresh. $\zeta_{\min}$ & $2.5$ \\
\hline
\end{tabular}
\end{table}

\subsection{Synchronization Accuracy}
\label{subsec:sync_results}
Table~\ref{tab:sync_rmse} shows that the correction removes nearly all injected timing and frequency errors. The timing error drops from the $15$\,ns TA jitter to below $1$\,ns and stays small over all $S$ occasions, so the cumulative sum does not drift. The receiver Doppler reference also vanishes, leaving a bias far below one Doppler bin, and the remaining phase error barely affects the coherent integration. Only a delay bias of tens of nanoseconds remains, which the delay gates absorb.

\begin{table}[t]
\centering
\caption{Synchronization accuracy averaged over $140$ trials.}
\label{tab:sync_rmse}
\footnotesize
\setlength{\tabcolsep}{4.0pt}
\begin{tabular}{l c c c c}
\hline
Metric & UE 1 & UE 2 & UE 3 & UE 4 \\
\hline
Timing RMSE [ns] & $0.88$ & $0.78$ & $0.80$ & $0.65$ \\
Phase RMSE [$^\circ$] & $0.57$ & $2.08$ & $1.02$ & $2.51$ \\
Residual delay bias $b_{k,\tau}$ [ns] & $-31.8$ & $+35.7$ & $-33.2$ & $+9.1$ \\
Doppler bias RMS $b_{k,\nu}$ [mHz] & $8.2$ & $42.5$ & $12.6$ & $62.3$ \\
\hline
\end{tabular}
\end{table}

\subsection{Detection and Localization Performance}
\label{subsec:detection_results}

We compare the proposed pipeline with four baselines and one ablation on the same synchronized and clutter-suppressed channels. The \emph{detect-then-fuse} (DTF) baseline follows the cooperative framework of~\cite{tang2025cooperative}, with peak picking on each link in place of its tensor decomposition, and clusters the peaks of all UEs by direction. The other baselines are CFAR with grid fusion (CFAR-GF)~\cite{liang2026uav}, multi-BS feature fusion (MBFF)~\cite{yan2026asynchronous}, and single-frame MUSIC with IAA (MUSIC-IAA)~\cite{he2026uplink}. The \emph{sum-fusion} ablation keeps our search but fuses the raw $\Lambda_k$ without the range normalization in Eq.~\eqref{eq:zeta}.

\begin{figure}[t]
    \centering
\begingroup
\definecolor{cOur}{RGB}{204,128,82}    
\definecolor{cOurL}{RGB}{240,214,192}  
\definecolor{cBlue}{RGB}{88,126,168}   
\definecolor{cGray}{RGB}{128,128,128}
\definecolor{cPur}{RGB}{138,114,166}   
\definecolor{cPink}{RGB}{186,104,138}
\definecolor{cPinkL}{RGB}{246,234,239}
\pgfplotsset{
  base/.style={
    tick label style={font=\scriptsize},
    label style={font=\scriptsize},
    axis line style={black!40, line width=0.4pt},
    axis lines=box, tick pos=left, tick align=outside,
    major tick length=1.8pt, every tick/.style={black!40},
    grid style={black!12, dashed, line width=0.3pt},
    legend style={font=\scriptsize, draw=black!25, fill=white, rounded corners=1.5pt,
      line width=0.3pt, inner xsep=2pt, inner ysep=1pt, row sep=-2.5pt, cells={anchor=west}},
  },
  bUse/.style={fill=cOur, draw=cOur, line width=0.45pt},
  bNear/.style={fill=cOurL, draw=cOur, line width=0.45pt},
  bFalse/.style={fill=cBlue!12, draw=cBlue, line width=0.45pt,
    postaction={pattern=north east lines, pattern color=cBlue!65}},
  bNone/.style={fill=black!8, draw=black!45, line width=0.45pt},
}
\tikzset{
  vlab/.style={font=\fontsize{5}{5}\selectfont, inner sep=0pt},
  vlabH/.style={font=\fontsize{5}{5}\selectfont, inner xsep=0.6pt, inner ysep=0.3pt, fill=white, fill opacity=0.9, text opacity=1},
}
\begin{tikzpicture}[font=\scriptsize]
\begin{axis}[
  name=A, base, scale only axis, width=2.9cm, height=2.2cm,
  xbar stacked, bar width=7.2pt, xmin=0, xmax=100, ymin=0.45, ymax=6.55,
  ytick={1,...,6}, ytick style={draw=none},
  yticklabels={MUSIC-IAA~\cite{he2026uplink},MBFF~\cite{yan2026asynchronous},CFAR-GF~\cite{liang2026uav},Detect-then-fuse~\cite{tang2025cooperative},Sum-fusion,\textbf{Proposed}},
  yticklabel style={font=\fontsize{6.5}{7}\selectfont, xshift=1pt},
  xtick={0,25,50,75,100}, xticklabels={0,25,50,75,100}, xmajorgrids,
  xlabel={(a) Trials [\%]}, xlabel style={yshift=3pt},
  legend style={at={(-0.2,1.04)}, anchor=south west, draw=none, fill=none,
    legend columns=4, column sep=2pt, inner sep=0pt},
  legend image code/.code={\draw[#1] (0cm,-0.08cm) rectangle (0.25cm,0.08cm);},
  clip=false,
]
\addplot[bUse]   coordinates {(0,1) (0,2) (1.4,3) (2.9,4) (7.1,5) (50.7,6)};
\addplot[bNear]  coordinates {(0,1) (0,2) (0,3) (0,4) (0,5) (7.1,6)};
\addplot[bFalse] coordinates {(100,1) (100,2) (98.6,3) (97.1,4) (65.7,5) (26.4,6)};
\addplot[bNone]  coordinates {(0,1) (0,2) (0,3) (0,4) (27.1,5) (15.7,6)};
\legend{within 10\,m, 10--18\,m, false decl., no decl.}
\end{axis}
\begin{axis}[
  at={(A.south west)}, anchor=south west, scale only axis, width=2.9cm, height=2.2cm,
  xmin=0, xmax=100, ymin=0.45, ymax=6.55, hide axis, clip=false,
]
\node[vlab, text=white, font=\fontsize{5}{5}\selectfont\bfseries] at (axis cs:25.35,6) {50.7};
\node[vlab, text=cOur!75!black, font=\fontsize{4.3}{4.3}\selectfont] at (axis cs:54.25,6) {7.1};
\node[vlabH, text=cBlue!85!black] at (axis cs:71.0,6) {26.4};
\node[vlab, text=black!60] at (axis cs:92.15,6) {15.7};
\node[vlab, text=white, font=\fontsize{4.3}{4.3}\selectfont] at (axis cs:3.55,5) {7.1};
\node[vlabH, text=cBlue!85!black] at (axis cs:39.95,5) {65.7};
\node[vlab, text=black!60] at (axis cs:86.45,5) {27.1};
\node[vlabH, text=cBlue!85!black] at (axis cs:51.45,4) {97.1};
\node[vlabH, text=cBlue!85!black] at (axis cs:50.7,3) {98.6};
\node[vlabH, text=cBlue!85!black] at (axis cs:50,2) {100};
\node[vlabH, text=cBlue!85!black] at (axis cs:50,1) {100};
\end{axis}
\begin{axis}[
  name=B, at={(A.east)}, anchor=west, xshift=12pt,
  base, scale only axis, width=2.32cm, height=2.2cm,
  xmode=log, xmin=1.5, xmax=400, ymin=0, ymax=1,
  xtick={2,10,100}, xticklabels={2,10,100}, minor xtick={3,30,300},
  ytick={0,0.5,1}, yticklabels={0,0.5,1}, minor ytick={0.25,0.75}, grid=major,
  ytick pos=right, yticklabel pos=right, ylabel near ticks,
  xlabel={(b) 3D position error [m]}, ylabel={CDF},
  xlabel style={yshift=3pt}, ylabel style={yshift=-2pt},
  legend style={at={(0.03,0.97)}, anchor=north west, font=\fontsize{5.5}{6}\selectfont,
    draw=black!30, fill=white, fill opacity=0.75, draw opacity=1, text opacity=1, sharp corners,
    inner xsep=1pt, inner ysep=1pt, row sep=-2pt},
  legend image post style={fill opacity=1, draw opacity=1},
  legend image code/.code={\draw[#1, mark repeat=1, mark phase=1, mark indices={2}] plot coordinates {(0cm,0cm) (0.12cm,0cm) (0.24cm,0cm)};},
  clip=true,
]
\fill[cPinkL] (axis cs:1.5,0) rectangle (axis cs:10,1);
\draw[cPink!80, densely dashed, line width=0.5pt] (axis cs:10,0) -- (axis cs:10,1);
\node[anchor=north west, text=cPink, font=\scriptsize, inner sep=1.5pt] at (axis cs:10,0.30) {10\,m};
\addplot[cOur, line width=1.1pt, const plot, mark=*, mark size=1.3pt, mark repeat=40, mark phase=20]
  table[x=x,y=y] {figure/fig4_cdf_prop.dat};
\addplot[cPur, line width=0.7pt, densely dotted, const plot, mark=diamond*, mark size=1.4pt, mark repeat=40, mark phase=10, mark options={solid}]
  table[x=x,y=y] {figure/fig4_cdf_ns.dat};
\addplot[cBlue, line width=0.7pt, const plot, mark=triangle*, mark size=1.4pt, mark repeat=40, mark phase=30]
  table[x=x,y=y] {figure/fig4_cdf_dtf.dat};
\addplot[cGray, line width=0.7pt, dash dot, const plot, mark=x, mark size=1.6pt, mark repeat=40, mark phase=25, mark options={solid}]
  table[x=x,y=y] {figure/fig4_cdf_liang.dat};
\legend{Proposed, Sum-fusion, Detect-then-fuse~\cite{tang2025cooperative}, CFAR-GF~\cite{liang2026uav}}
\end{axis}
\begin{axis}[
  at={(A.south west)}, anchor=north west, yshift=-27pt,
  scale only axis, width=5.643cm, height=1.7cm, xmin=0.4, xmax=14.6, ymin=0, ymax=146,
  hide axis, clip=false,
]
\addplot[draw=none, fill=cOur!7, forget plot] coordinates {(0.4,146) (9.5,146)} \closedcycle;
\end{axis}
\begin{axis}[
  name=C, at={(A.south west)}, anchor=north west, yshift=-27pt,
  base, scale only axis, width=5.643cm, height=1.7cm,
  ybar stacked, bar width=7.5pt, ymin=0, ymax=146, xmin=0.4, xmax=14.6,
  xtick={1,...,14}, xticklabels={36,42,37,45,48,41,39,38,44,40,49,47,46,43},
  xtick style={draw=none},
  ytick={0,50,100}, minor ytick={25,75}, ymajorgrids,
  xlabel={(c) Static-clutter realization}, ylabel={Trials [\%]},
  xlabel style={yshift=3pt}, ylabel style={yshift=-7pt},
  clip=false,
]
\addplot[bUse]   coordinates {(1,90) (2,90) (3,80) (4,80) (5,70) (6,60) (7,50) (8,50) (9,50) (10,30) (11,30) (12,10) (13,10) (14,10)};
\addplot[bNear]  coordinates {(1,0) (2,0) (3,0) (4,0) (5,0) (6,0) (7,30) (8,0) (9,0) (10,10) (11,0) (12,30) (13,20) (14,10)};
\addplot[bFalse] coordinates {(1,0) (2,10) (3,10) (4,0) (5,30) (6,20) (7,20) (8,30) (9,50) (10,20) (11,50) (12,50) (13,30) (14,50)};
\addplot[bNone]  coordinates {(1,10) (2,0) (3,10) (4,20) (5,0) (6,20) (7,0) (8,20) (9,0) (10,40) (11,20) (12,10) (13,40) (14,30)};
\draw[cPink, densely dashed, line width=0.6pt] (axis cs:0.4,50.7) -- (axis cs:14.6,50.7);
\node[anchor=west, text=cPink, font=\scriptsize, inner sep=1pt, align=left] at (axis cs:14.6,50.7) {mean\\[-2pt]50.7\%};
\draw[black!30, densely dashed, line width=0.4pt] (axis cs:9.5,0) -- (axis cs:9.5,146);
\node[anchor=north west, inner sep=1.5pt, align=left, font=\fontsize{6.5}{7}\selectfont, text=cOur!70!black]
  at (axis cs:0.4,146) {9 realizations with $\geq$50\% useful trials:\\91.1\% declared, 75.6\% of them $\leq$10\,m};
\node[anchor=north west, inner sep=1.5pt, align=left, font=\fontsize{6.5}{7}\selectfont, text=black!60]
  at (axis cs:9.5,146) {5 realizations:\\10--30\% useful};
\end{axis}
\end{tikzpicture}%
\endgroup%
%
    \caption{Results on the $140$ trials. (a) Detection rates. (b) CDF of the 3D position error. (c) Proposed pipeline for each static-clutter realization.}
    \label{fig:cdf}
\end{figure}

Fig.~\ref{fig:cdf}(a) compares the detection rates. The proposed pipeline localizes the UAV within $10$\,m in $50.7\%$ of the trials, with $7.1\%$ near misses at $10$--$18$\,m, whereas no baseline exceeds $3\%$. The baselines declare a target in every trial, but nearly all declarations are false. Their detectors work on each link separately and keep the strongest peaks of each UE, which belong to clutter when the echo is $44$--$49$\,dB below the LOS, so fusing these peaks afterwards yields clutter combinations. Sum-fusion searches the same state space, yet most of its declarations are still false, so the range normalization is what lets the weak echo stand out.
Fig.~\ref{fig:cdf}(b) shows the same contrast in the position error. The proposed CDF rises steeply below $10$\,m, whereas the baseline errors spread over tens to hundreds of meters. The useful detections have a median 3D error of $4.84$\,m, limited mainly by the angular resolution of the BS array. Along the BS ray, the median range error is only $2.5$\,m, because the UE delays fix $\hat\rho$ in Eq.~\eqref{eq:rho_closed}. Across the ray, the azimuth and elevation RMSEs of $2.4^\circ$ and $5.1^\circ$ of the $4\times4$ array give about $2$\,m and $5$\,m at the $53$\,m UAV range, and the ground UEs add little elevation diversity. We therefore expect a larger BS array to improve localization more than a wider bandwidth.
Fig.~\ref{fig:cdf}(c) shows that the static-clutter layout dominates the performance. Most realizations localize the UAV in at least half of their trials, and the failures concentrate in a few layouts. Since only the clutter changes, these failures come from clutter paths rather than from a weak echo.

A failed trial declares nothing or a ghost state. A ghost is formed by clutter paths of different UEs whose delays and angles agree by chance, and it does not lie on any physical scatterer. Cross-UE consistency can thus also favor a ghost. The failure lies in the ranking, not in the search: the coarse search keeps a candidate within $10$\,m of the UAV in $117$ of $140$ trials, but after local refinement a ghost often outscores it. A higher declaration threshold does not help, as ghosts currently score like the UAV. Separating them needs a score that ranks the UAV higher, or evidence such as rotor micro-Doppler or track continuity.

\subsection{Robustness Analysis}
\label{subsec:nlos}
To assess the effect of NLOS references and UE position errors, Fig.~\ref{fig:robust} compares the useful rate in (a) and the median error in (b) with the nominal values. Resampling the receiver noise shows the Monte Carlo variation, and UE position errors of a few meters change the rate even less, so the delay constraints tolerate them.
NLOS has a larger effect. When the LOS of UE~1, or of UEs~1 and~2, is blocked, a static reflection longer by $\Delta_{\mathrm{ex}}$ becomes the reference. Synchronization still works, with $0.86$\,ns timing RMSE on the blocked UE against $0.88$\,ns with LOS (Table~\ref{tab:sync_rmse}), but all delays of that UE shift by $\Delta_{\mathrm{ex}}/c$. A $3$\,m excess shifts them by $10$\,ns, inside the $35$--$45$\,ns delay gates, and costs only part of the detections. A $60$\,m excess shifts them by $200$\,ns and removes the UE from the fusion, so the three remaining UEs must all support the UAV, and two blocked UEs leave fewer than $K_{\min}=3$, hence the sharp drop in (a) and the larger errors in (b). The loss thus comes from the delay gates rather than the synchronization, and a refinement that tolerates a shifted UE may recover it. Each NLOS case uses one synthetic reflection with a fixed excess length, so ray-traced channels are needed for a realistic assessment. Several UAVs, which would give separate peaks of $T(\boldsymbol{\theta})$, and a hovering UAV, whose body echo the mean subtraction removes, are not evaluated.

\begin{figure}[t]
    \centering
\begingroup
\definecolor{cOur}{RGB}{204,128,82}
\definecolor{cBlue}{RGB}{88,126,168}
\definecolor{cPink}{RGB}{186,104,138}
\pgfplotsset{
  rob/.style={
    scale only axis, height=1.6cm, ymin=0.45, ymax=6.55,
    tick label style={font=\scriptsize}, label style={font=\scriptsize},
    axis line style={black!40, line width=0.4pt}, axis lines=box,
    tick pos=left, tick align=outside, major tick length=1.8pt, every tick/.style={black!40},
    ytick={6,5,4,3,2,1}, ytick style={draw=none},
    xmajorgrids, grid style={black!12, dashed, line width=0.3pt},
    xlabel style={yshift=3pt}, clip=false,
    nodes near coords, point meta=explicit symbolic,
    nodes near coords style={font=\scriptsize, text=black!70, anchor=west, xshift=1pt,
      fill=white, inner xsep=1pt, inner ysep=0.3pt},
  },
}
\begin{tikzpicture}[font=\scriptsize]
\begin{axis}[
  name=L, rob, width=2.9cm, xbar, bar shift=0pt, bar width=5.4pt, xmin=0, xmax=66,
  yticklabels={Nominal, Resampled noise, UE pos.\ err.\ $\sigma{=}3$\,m,
    NLOS UE\,1 $\Delta_{\mathrm{ex}}{=}3$\,m, NLOS UE\,1 $\Delta_{\mathrm{ex}}{=}60$\,m,
    NLOS UE\,1--2 $\Delta_{\mathrm{ex}}{=}60$\,m},
  xtick={0,20,40,60}, xlabel={(a) Useful detections [\%]},
]
\draw[cPink, densely dashed, line width=0.5pt] (axis cs:50.7,0.45) -- (axis cs:50.7,6.55);
\addplot[fill=cOur, draw=cOur, line width=0.45pt] coordinates {(50.7,6) [50.7]};
\addplot[fill=cBlue!25, draw=cBlue, line width=0.45pt] coordinates {
  (43.6,5) [43.6] (49.3,4) [49.3] (38.6,3) [38.6] (16.4,2) [16.4] (5.0,1) [5.0]};
\end{axis}
\begin{axis}[
  name=R, at={(L.east)}, anchor=west, xshift=10pt,
  rob, width=1.7cm, xbar, bar shift=0pt, bar width=5.4pt, xmin=0, xmax=12, yticklabels={,,},
  xtick={0,4,8}, xlabel={(b) Median err.\ [m]},
]
\draw[cPink, densely dashed, line width=0.5pt] (axis cs:4.84,0.45) -- (axis cs:4.84,6.55);
\addplot[fill=cOur, draw=cOur, line width=0.45pt] coordinates {(4.84,6) [4.84]};
\addplot[fill=cBlue!25, draw=cBlue, line width=0.45pt] coordinates {
  (4.75,5) [4.75] (4.85,4) [4.85] (5.09,3) [5.09] (8.12,2) [8.12] (7.16,1) [7.16]};
\end{axis}
\end{tikzpicture}%
\endgroup%
%
    \caption{Robustness of the proposed pipeline. (a) Useful detection rate. (b) Median 3D error of the useful detections.}
    \label{fig:robust}
\end{figure}

\section{Conclusion}
\label{sec:conclusion}
This paper presented the first practical passive UAV detection and localization framework that reuses standard 5G NR uplink SRS from multiple mobile UEs. LOS-referenced synchronization removes the timing, phase, and gain distortions, and geometry-coupled fusion turns weak uplink echoes into useful localization in simulation, where \emph{detect-then-fuse} baselines fail. Future work will address ghost suppression, UAV tracking, and validation on measured channels.

\bibliographystyle{IEEEtran}
\bibliography{IEEEabrv,ref}

\end{document}